\documentclass[aps,prc,preprint,showpacs,amsmath,amssymb,nofootinbib]{revtex4}
\usepackage {lscape}
\usepackage{graphicx}
\usepackage{amsmath}
\usepackage{color}
\usepackage{mathrsfs}

\usepackage{epsfig,colordvi}

\usepackage{graphicx}
\usepackage{dcolumn}
\usepackage{bm}
\usepackage{epsfig}
\usepackage{array}
\usepackage{amsmath}
\usepackage[final]{pdfpages}
\usepackage{pdfpages}

\usepackage{amsfonts}
\usepackage{amssymb}
\usepackage{amsmath}
\usepackage{amsxtra}

\usepackage[bookmarks=true]{hyperref}
\usepackage{color}

  \hypersetup{
   colorlinks=true,
   pdfborder={0 0 0},
   citecolor=blue,
   linkcolor=blue,
   urlcolor=blue
 }

\begin{document}

\title{Ab initio no-core shell model study of $^{18-23}$O and $^{18-24}$F isotopes}

\author{ Archana Saxena\footnote{archanasaxena777@gmail.com}  and Praveen C. Srivastava\footnote{Corresponding author: praveen.srivastava@ph.iitr.ac.in}}

\address{Department of Physics, Indian Institute of Technology Roorkee, Roorkee
247 667, India}

\date{\hfill \today}

\begin{abstract}

In the present work, we have done a  comprehensive study of  low-lying energy spectrum  for oxygen  and fluorine chains using $ab~initio$ no core shell model. 
We have used inside nonlocal outside Yukawa (INOY) potential, which is a two body interaction but also has the effect of three body forces by short range and nonlocal character. 
Also, we have performed calculations with N3LO and N2LOopt interactions and compared corresponding results with the experimental data and phenomenological USDB interaction. 
We have reached up to $N_{max}$=6 for $^{18-21}$O and $^{18-19}$F, $N_{max}$=4
for other oxygen and fluorine isotopes, respectively.
We have also discussed the binding energy of oxygen  and fluorine chains.
Over binding in the ground state (g.s.) energy in neutron rich oxygen isotopes is observed in our largest model space calculations. 
\end{abstract}

\maketitle

\section{\bf{Introduction}}
Recent developments in computing power have made it possible to study the nuclei beyond $p$ shell using no core shell model (NCSM) approach
\cite{Barrett,Navratil1,Navratil2,Navratil3,Forssen1,archana,soma,andray1,andray2}.
The NCSM is a basic tool for explaining the nuclear structure and nuclear bound state problems,
which uses nuclear interactions from first principles. Now, calculations are also available with the continuum effect in NCSM for explaining unbound states, scattering and nuclear reactions. Recently, NCSM calculations with continuum effect for $^{11}$Be have been performed, the parity inversion problem is solved for $^{11}$Be  using 
NCSMC approach  \cite{Baroni1,Baroni2, Navratil} with chiral interaction N$^2$LO$_{SAT}$ \cite{Calci}. 
$Ab~initio$ approaches give an opportunity to calculate the electromagnetic properties of exotic nuclei more precisely  
 \cite{Forssen}. The NCSM calculations  have various applications in 
nuclear structure as well as in nuclear reaction physics. New physics is coming out, e.g. $N$=8 magic number is no more magic number when we 
increase neutron and proton ratio. 
There is another method known as no core shell model with core constructed for the heavier nuclei \cite{Barrett,Dikmen,nadya,em,GT,hyperfine,21Mg,28Mg}.
To understand the details of fully open shell medium mass nuclei from first principle is still a challenge.
Research for open shell nuclei is still going on using many $ab~initio$ approaches \cite{Stroberg, Roth}. 
The study of neutron rich nuclei from first principles is an interesting topic nowadays. Exact location of drip line 
in neutron rich oxygen isotopes can be explained by $ab~initio$ approaches using NN+3N(full) chiral interactions 
where the role of 
3N forces is very important \cite{Otsuka}.  The no core shell model with perturbation approach (NCSM-PT) results for low-lying states in $^{18-20}$O isotopes are reported in Ref. \cite{Roth}. 

 In the present work our aim is to systematically study the low-lying energy spectrum (positive parity) for oxygen ($^{18-23}$O) and fluorine ($^{19-24}$F) chains using $ab~initio$ no core shell model. 
First time we have reached up to $N_{max}$ = 6 calculations for $^{18-21}$O and $^{18,19}$F isotopes.

The paper is organized as follows:
In Secs. II and III, the NCSM and its formalism are given. The effective interactions used in calculations are described in Sec. IV. Results and discussions are reported in Sec. V. Neutron drip line in oxygen isotopes discussed in Sec. VI. Finally the paper is concluded in Sec. VII.

\section{\bf{$Ab~initio$} No Core Shell Model}
$Ab~initio$ no core shell model \cite{Navratil1,Navratil2} is a many body approach which is used to solve full $A$- body Hamiltonian.  No core is assumed in this model, all the nucleons are treated as active non relativistic point 
like particles interacted by realistic $NN$ or $NN+NNN$ interactions which fit Nijmegan $NN$ phase shifts  up to a energy 350 MeV with a very high precision. However,
in the shell model, we assume a core, so, the core particles are considered to be inactive. For NCSM, we choose Harmonic Oscillator (HO) basis for the calculations because in this we can use second quantization 
formalism and our system remains translationally invariant. The $NN$ potentials which generate short range correlations cannot be accommodated in HO basis 
for e.g. Argonne V18(AV18) \cite{Wiringa}, CD-Bonn 2000 \cite{Machleidt} and INOY  \cite{{Doleschall1,Doleschall2}} potentials.
Chiral potentials \cite{Entem} also produce short range 
correlations up to some extent. To accommodate these correlations 
and speeding up the convergence with the expansion of basis, we need an effective interaction from starting realistic $NN$ and $NN+NNN$ interactions.
Effective interactions 
can be generated by using unitary transformation. Many soft potentials (which do not generate short range correlations) are also available nowadays for e.g. $V_{lowk}$ \cite{Bogner1}, SRG $NN$ \cite{Bogner2} etc. In the present work, we have also compared our NCSM results with phenomenological USDB effective interaction
\cite{usdb}.

\section{\bf{Formalism }}

We have to solve A-body Schr\"{o}dinger equation. The starting Hamiltonian is given as-
\begin{eqnarray}
\label{eq:(1)}
H_{A} = T_{rel} + V  = \frac{1}{A} \sum_{i< j}^{A} \frac{({\vec p_i - \vec p_j})^2}{2m} 
 +\sum_{i<j}^A V_{NN,ij} 
\end{eqnarray}

where $m$ is the mass of the nucleon and $A$ is the number of nucleons ($i.e.,$ the mass number). The first term is relative kinetic energy defined in terms of momenta of each nucleon ($p_{i,j}= 1,2,3...A$), second term is nucleon-nucleon $NN$ interaction having nuclear and Coulomb part both. 
We modify Eq. \ref{eq:(1)}, by adding the HO center-of-mass term
 $H_{CM}= T_{CM}+ U_{CM}$, where $U_{CM}=\frac{1}{2}Am{\Omega}^2 {\vec R}^2$,
$\vec R=1/A\sum_{i=1}^A {\vec r_{i}}$ and $\Omega$ is the HO frequency.

\begin{eqnarray}
\label{eq:(2)}
H^{\Omega}_{A} = H_{A} + H_{CM} =\sum_{i=1}^A h_{i} + \sum_{i< j}^A {V_{ij}}^{\Omega,A} \nonumber 
 = \sum_{i=1}^A \left[ \frac {{\vec p_i } ^2} {2 m} + \frac{1}{2}m {\Omega}^2{\vec r_i}^2 \right] \nonumber\\
 +\sum_{i<j}^A \left[ V_{NN,ij} - \frac{1}{2A}m {\Omega}^2 {({\vec r_i - \vec r_j})}^2 \right]. 
\end{eqnarray}

In the final calculations CM term will be subtracted out. 
The new Hamiltonian breaks into two parts, one body and two body parts. Now, we divide our basis space into two parts, one in which we do the calculations,
called $P$ model space, which is finite and second one is Q($=1-P$) model space, which is excluded space. 
Since the Harmonic-oscillator basis is infinite, we make a truncation in the $Q$ space (see Ref. 
\cite{NavratilPRC54} for more details).

These operators satisfy the conditions $P^2=P$, $Q^2=Q$, $PQ=0$, and 
$P+Q=1$. We introduce an operator, $\omega$, which is a transformation operator of similarity transformation of the Hamiltonian $e^{-\omega}He^{\omega}$, satisfies the condition $\omega= Q \omega P$. This operator acts as a mapping between the P and Q spaces. 
The eigenstates of initial Hamiltonian can be denoted by $|k\rangle$ with eigenvalues $E_{k}$.

 \begin{equation}
 H |k\rangle=E_{k}|k\rangle.
 \label{H1}
 \end{equation}
 
The basis states of P and Q spaces are expressed as $|\alpha_{P}\rangle$ and $|\alpha_{Q}\rangle$. The $|\alpha_{P}\rangle$ belongs to $|\alpha_{Q}\rangle$ 
and follow the relation $Q e^{-\omega}He^{\omega} P=0$. The $|\alpha_{Q}\rangle$ states can be expressed in terms of $|\alpha_{P}\rangle$ states 
with the help of $\omega$ operator\cite{NavratilPRC62},

\begin{equation}
\langle \alpha_{Q}|k\rangle=\sum_{\alpha_{P}} \langle \alpha_{Q}|\omega|\alpha_{P}  \rangle  \langle\alpha_{P}|k\rangle.
\label{ak}
\end{equation}

The operator  $\omega$ can be find out from Eq. \ref{ak}. Let $d_{P}$ is the dimension of model space P and $\mathscr{K}$ which is a set of the $d_{P}$
eigen vector vectors which satisfies the relation \ref{ak}. The $d_{P} \times d_{P}$ which is a matrix having matrix elements
$\langle \alpha_{P}|k\rangle$ for $|k\rangle \in \mathscr{K}$ is invertible, under this condition the operator $\omega$,

\begin{equation}
\langle \alpha_{Q}|\omega|\alpha_{P}\rangle = \sum_{k \in \mathscr{K}} \langle \alpha_{Q}|k\rangle  \langle \tilde{k}|\alpha_{P}\rangle,
\label{solvew}
\end{equation}

where tilde denotes the matrix elements of inverse matrix $\langle\alpha_{P}|k\rangle$,
e.g., \\ $\sum_{\alpha_{P}} \langle \tilde{k}|\alpha_{P}\rangle \langle \alpha_{P}|k'\rangle = \delta_{k,k'}$, for $k,k' \in \mathscr{K}$.

For model space P, the Hermitian effective Hamiltonian is given by 
\begin{equation}
\bar{H}_{eff}=[P(1+ \omega\dagger\omega)P]^{1/2}PH(P+Q\omega P) [P(1+ \omega\dagger\omega)P]^{-1/2}.
\end{equation}
This Hamiltonian can be rewritten by using properties of $\omega$ operator

\begin{equation}
\bar{H}_{eff}=[P(1+ \omega \dagger \omega)P]^{-1/2}(P+P \omega \dagger Q)H(Q\omega P +P)[P(1+\omega\dagger\omega)P]^{-1/2}.
\label{Heff_1}
\end{equation}


We need effective interactions to employ in our model space because of short 
range repulsive nature of realistic $NN$ interactions. For this purpose, we perform a unitary transformation. 
We apply unitary transformation to Eq. \ref{eq:(2)} such that the model space $P$ and $Q$ decoupled to each other \cite{Suzuki1,Suzuki2}. 
After applying unitary transformation we get Hermitian effective
interaction $H_{eff}=Pe^{-S}H_{A}^\Omega e^{S} P$. $H_{eff}$ is an $A$-body operator. 
In our approximation the Hamiltonian is at two-body cluster level, for details, see Refs. \cite{Barrett,Forssen}. 
We have following variational parameters in the final Hamiltonian; first one is the harmonic oscillator 
frequency $\hbar\Omega$, the second one is nucleon number $A$ and third one is the size of the model space $N_{max}$, where $N_{max}$ is 
the parameter that measures the maximal allowed HO excitation energy 
above the unperturbed ground state. When $N_{max}\to \infty$ ,  $V_{ij,eff}^{\Omega,A} \to {V_{ij}}^{\Omega,A}$.
So, as we increase model space, the dependency on $\Omega$ decreases and we get the converge result.
In NCSM, effective interactions are translationally  invariant. Because of this we can decouple the c.m. components from all observables. 
Our effective Hamiltonian is in $A$-body space. To construct this, we subtract the  c.m. Hamiltonian and add the Lawson projection term \cite{Lawson} 
${\beta (H_{c.m.} - \frac{3}{2} \hbar \Omega )}$ to shift spurious states which comes because of c.m. excitations. 
In the present work we have taken $\beta$ = 10. The eigenvalues of physical states are independent of the $\beta$. Now the effective Hamiltonian [Eq. \ref{eq:3}] is given as:



\begin{eqnarray}
H_{A,eff}^{\Omega} = P \Bigg\{ \sum_{i=1}^A \Bigg[ \frac{({\vec p_i - \vec p_j})^2}{2mA} + \frac{m {\Omega}^2 }{2A}({\vec r_i-\vec r_j})^2  
 +V_{ij,eff}^{\Omega,A} \Bigg] 
+ \beta \left( H_{c.m.} - \frac{3}{2} \hbar \Omega \right)  \Bigg\} P. 
\label{eq:3}
\end{eqnarray}

Computationally, the eigenvalues of the many body Hamiltonian is not easy task because it involves a very huge matrix to  diagonalize.
 In the present paper we have used the pANTOINE  \cite{pantoine, Caurier1,Caurier2} shell model code which is adapted to NCSM \cite{Caurier3}. 
 The NCSM is a sparse matrix eigenvalue problem. 
The dimensions of the Hamiltonian matrix increases with  atomic number $A$ and increasing 
$N_{max}$. In the case of heavier neutron rich O and F isotopes, the dimensions are very large. Thus, we have performed calculations up to $N_{max}$ = 4 only. In our calculations, 
in the case of $^{21}$O, corresponding to $N_{max}$ = 6, dimension is $\sim$ 3.36 x 10$^9$ ;
for  $^{19}$F, corresponding to $N_{max}$ = 6, dimension is $\sim$ 1.35 x 10$^9$.

\section{\bf{ Effective $NN$ Interaction}}
In NCSM, as we increase model space size, solving many body problem becomes computationally hard.
We want to do calculations with $NN$ interactions in maximum model space size. Previously, the INOY interaction is used in NCSM calculations to find the binding energies, excited states of both parities, electromagnetic moments, and point-nucleon radii in Be isotopes \cite{Forssen}.

There are two reasons behind choosing  INOY potential, first one is that it contains three body effect through nonlocality in some partial waves so we get the effect of three body forces also without adding three body forces explicitly and
the second reason is that we get fast convergence for INOY interaction for a given $N_{max}$ in comparison to any other interaction. 
The nonlocality and three body forces are deeply related to each other \cite{Polyzou}. The
INOY potential is a nonlocal potential in a coordinate space. Actually it is a mixture of local and nonlocal parts, local Yukawa tail at
longer ranges ($>$ 3 fm) and 
nonlocal at short range. As we know nucleons have  internal structure, because of this a nonlocality character comes at short range (up to 1-1.5 fm). The
INOY $NN$ interaction ( set of $^{1}{S_0}$ and $^{3}{SD_1}$ $NN$ interactions ) is in coordinate space and it was constructed to see the effect on triton binding energy.
When we use the coordinate space, it is easy to handle the ranges of local and nonlocal parts explicitly. In coordinate space, because of the basic 
property (short range nature) of $NN$ interaction, it should vary as an exponential function at long ranges.

 The form of the INOY $NN$ interaction 
 \cite{Doleschall1,Doleschall2}  is given by

\begin{eqnarray}
V^{full}_{ll'}(r,r')= W_{ll'}(r,r')+ \delta (r-r') F^{cut}_{ll'}(r) V^{Yukawa}_{ll'}(r),
\end{eqnarray}

where,

\[ F^{cut}_{ll'}(r) =
  \begin{cases}
      1- e^{-[\alpha_{ll'}(r-R_{ll'})]^{2}} & for ~r\geq R_{ll'} \\
     0 & for ~r\leq R_{ll’}  \\
  \end{cases}
\]

$W_{ll'}(r,r')$ and $V^{Yukawa}_{ll'}(r)$ parts are the nonlocal and Yukawa tail (it is taken from AV18 potential \cite{Wiringa}), respectively. The $F^{cut}_{ll'}(r)$ is the cut off function for Yukawa tail.
The $\alpha_{ll'}$ and $R_{ll'}$ are the fixed parameters which are 1.0 fm$^{-1}$ and 2.0 fm, respectively. The nonlocal form $W_{ll'}(r,r')$ with parameters is given in Ref. \cite{Doleschall1}. These nonlocal INOY interactions are phenomenological and reproduces the 3$N$ binding energies without adding a 3$N$ force.


 Apart from INOY, in the present work we use the interactions N3LO \cite{Entem},  N$^{2}$LO$_{opt}$ \cite{PRL110192502} and USDB \cite{usdb}. 
The N3LO \cite{Entem} is nucleon-nucleon interaction at the next-to-next-to-next-to-leading order (fourth order) of chiral perturbation theory  (􏰋􏰲$\chi$PT)􏰌. In this charge dependence is included up to next-to-leading order of the isospin-violation scheme. 
In $\chi$PT, the $NN$ amplitude is uniquely determined by two classes of contributions: contact terms and pion-exchange diagrams.  At N3LO, there are 24 parameters  corresponding to 24 contact terms for the fitting of partial waves.
The N$^{2}$LO$_{opt}$ \cite{PRL110192502} is optimized nucleon-nucleon interaction from chiral effective 
field theory at next-to-next-to-leading order (NNLO). Three pion-nucleon ($\pi$$N$) couplings ($c_{1}$, $c_{3}$, $c_{4}$) and 11 partial wave contact parameters C and $\tilde{C}$ were optimized. 
N$^{2}$LO$_{opt}$ interaction yields very good agreement with binding energies and radii for $A$ = 3, 4 nuclei.
N$^{2}$LO$_{opt}$ \cite{PRL110192502} is a bare interaction.  
In the bare interaction, Okubo-Lee-Suzuki (OLS) \cite{Suzuki1,Suzuki2} transformation is not used. 
The USDB \cite{usdb}  interaction for shell model calculations with core  is  based on a renormalized G matrix with linear combinations of two-body matrix elements adjusted to fit a complete set of data for experimental binding energies and excitation energies for the $sd$-shell nuclei. This interaction was fitted by varying 56 linear combinations of parameters. 
This interaction is very successful to explain  nuclear structure and nuclear astrophyical properties of $sd$ shell nuclei.
   
The N3LO calculations are done at
 $\hbar\Omega$=14 MeV. 
Previously, the NCSM calculations are done using N3LO $NN$ interaction for $^{18}$F up to $N_{max}$=4 \cite{Barrett}. 
In the present calculations we have reached up to $N_{max}$=6 in case of $^{18,19}$F. The calculations with next-to-next
leading order (N2LOopt) interaction at $\hbar\Omega$=20 MeV have been also reported. 
In the present calculations, we also see the difference between the results using bare and all other interactions (N3LO, INOY and USDB). 

\begin{figure*}
\begin{center}
\includegraphics[width=8cm,height=7.5cm,clip]{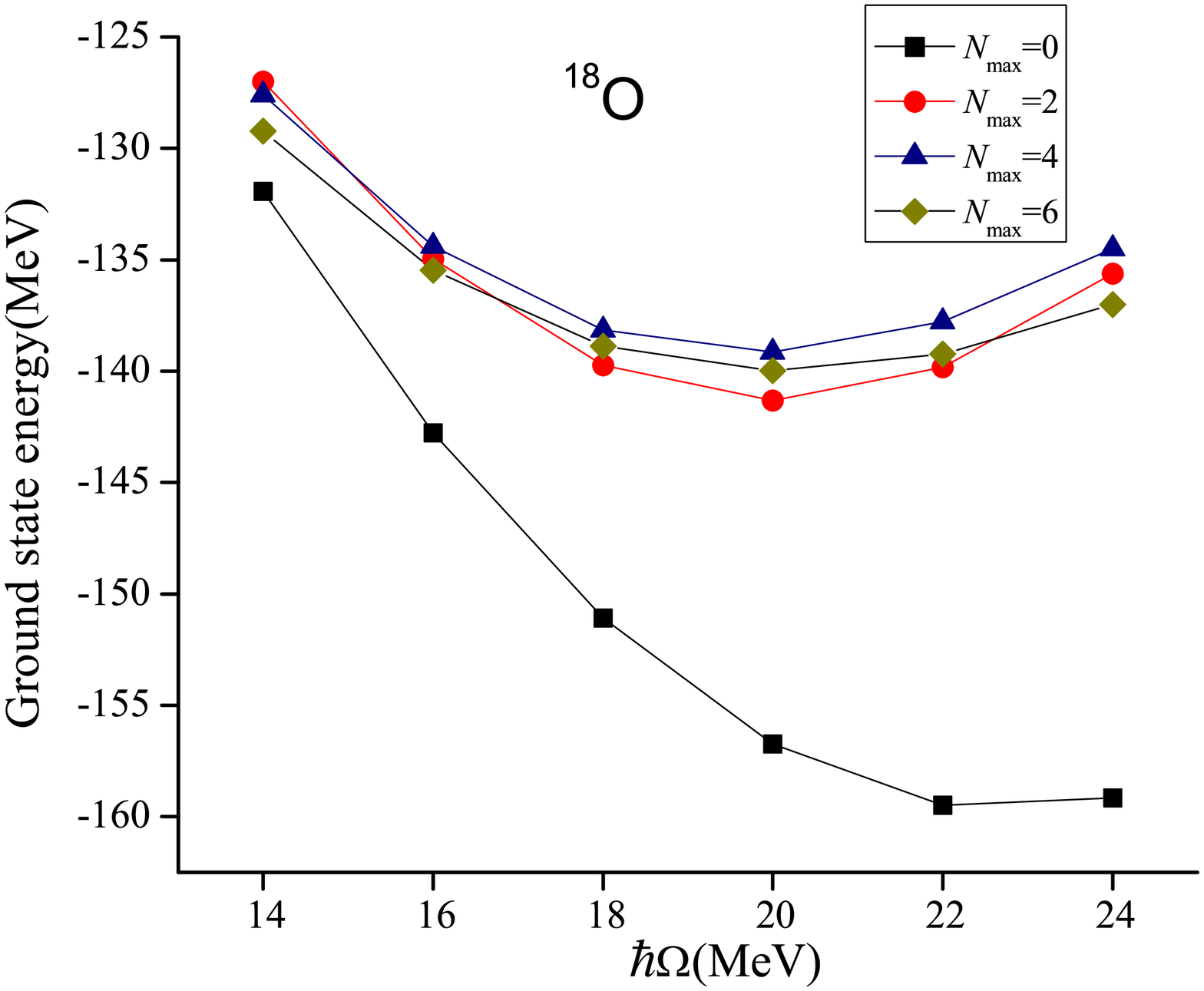}
\includegraphics[width=8cm,height=7.5cm,clip]{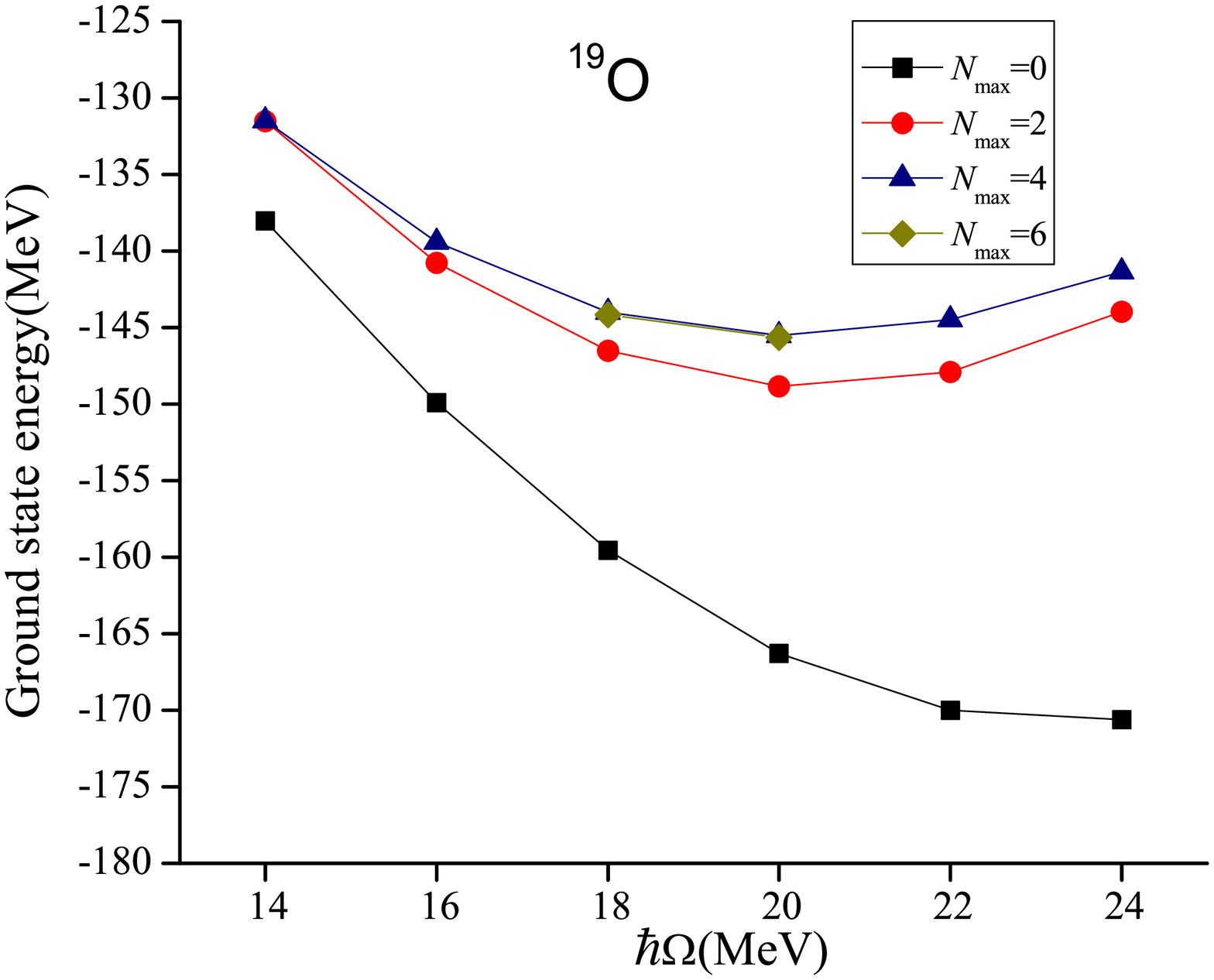}
\caption{\label{Oxygen_hw}The variation of g.s. energy with frequencies at different model spaces for $^{18,19}$O isotopes, similarly for other O isotopes.}
\end{center}  
\end{figure*}

\begin{figure*}
\includegraphics[width=16cm,height=7.5cm,clip]{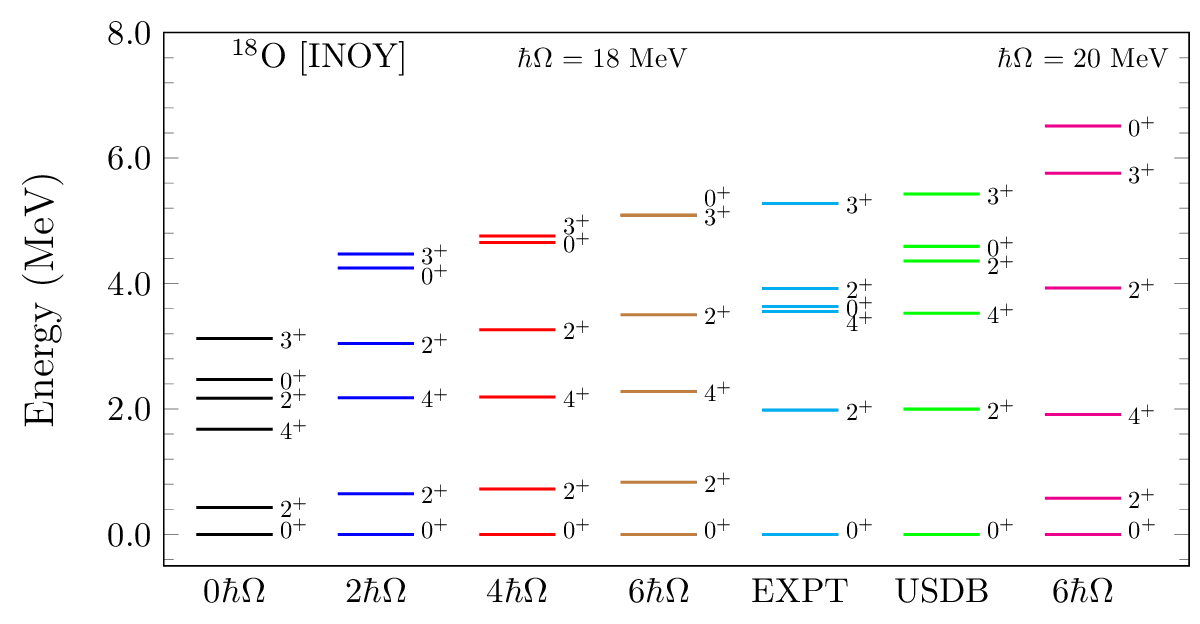}
\includegraphics[width=16cm,height=7.5cm,clip]{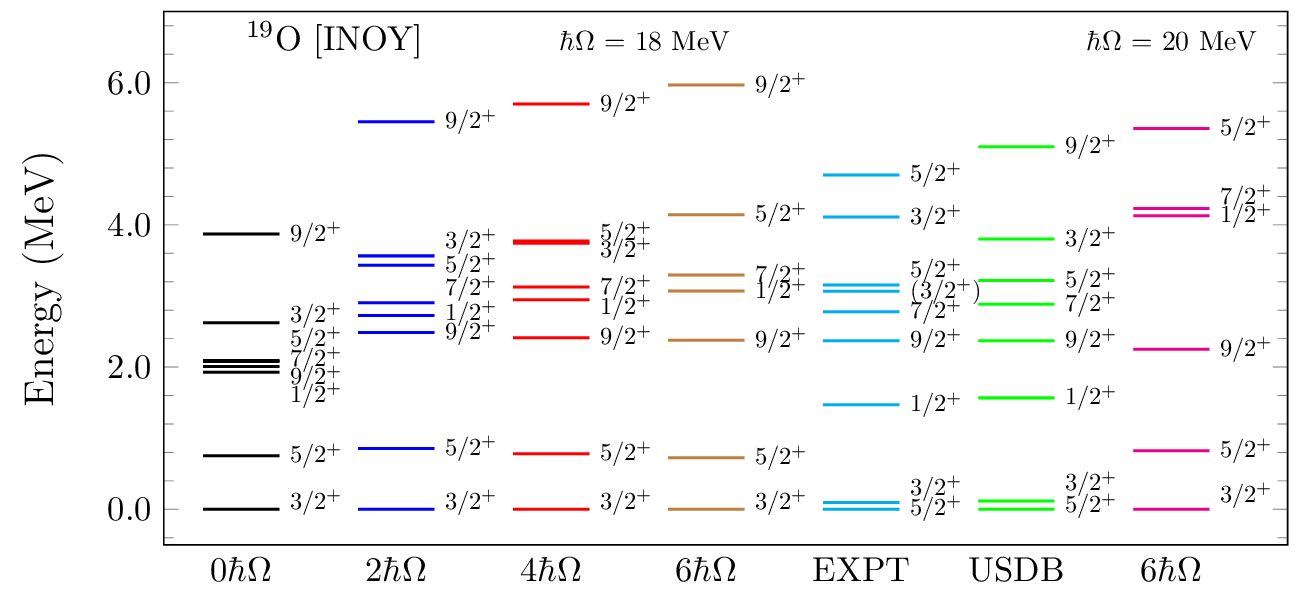}
\includegraphics[width=16cm,height=7.5cm,clip]{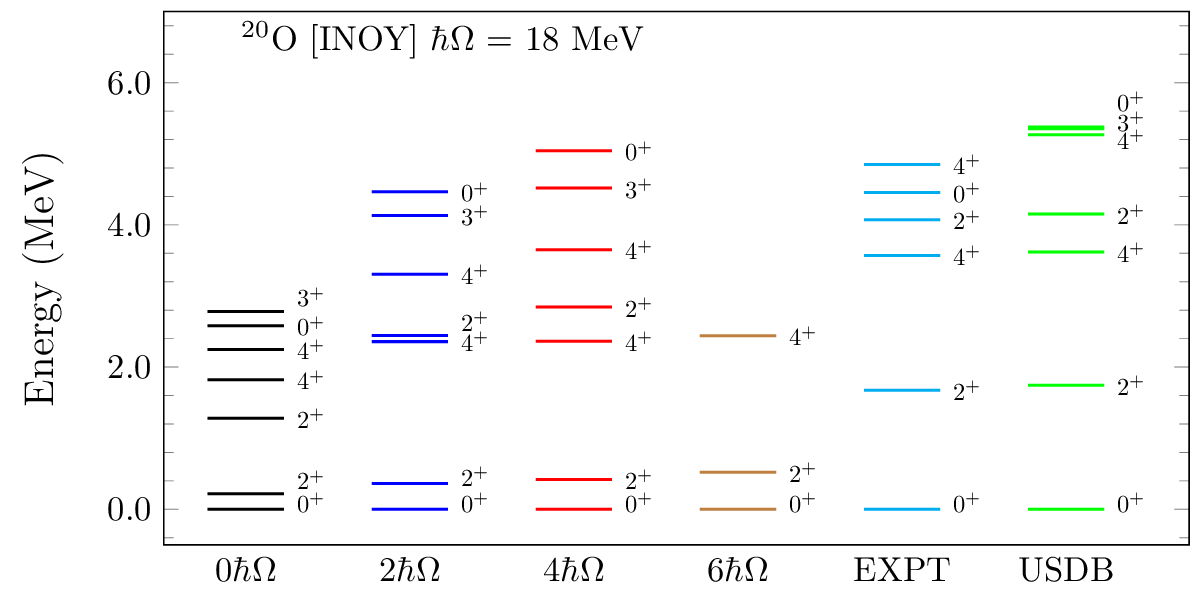}
\caption{\label{O_1820} The energy spectra of  $^{18-20}$O isotopes with INOY and USDB interactions.}
\end{figure*}

\begin{figure*}
\begin{center}
\includegraphics[width=12cm,height=7.0cm,clip]{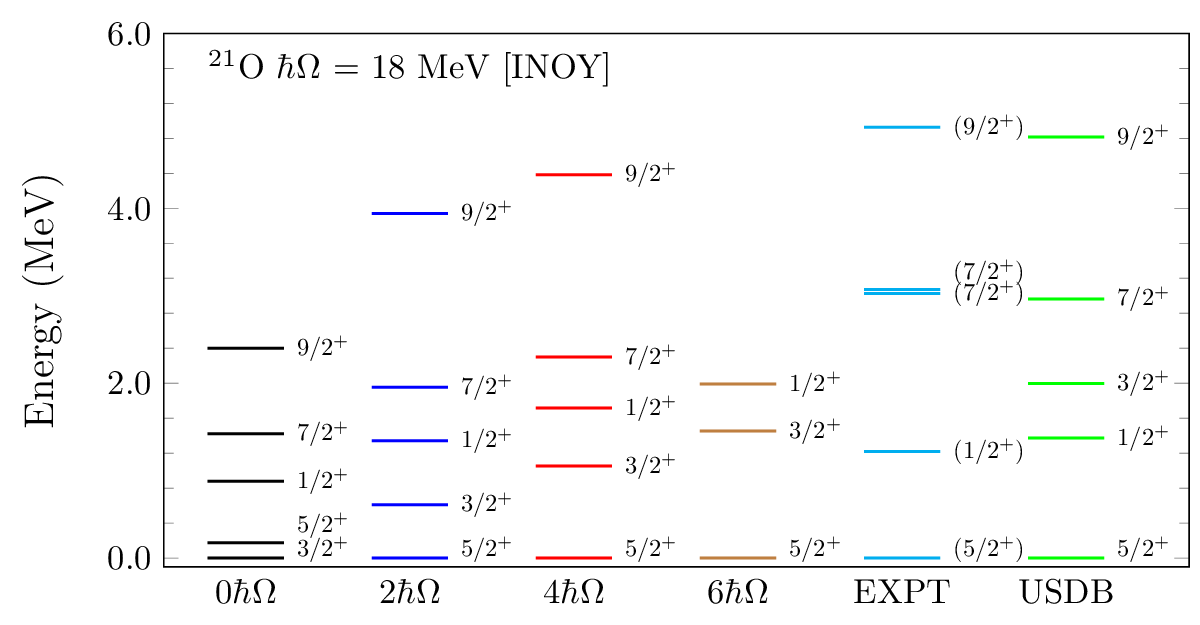}
\includegraphics[width=8cm,height=7.0cm,clip]{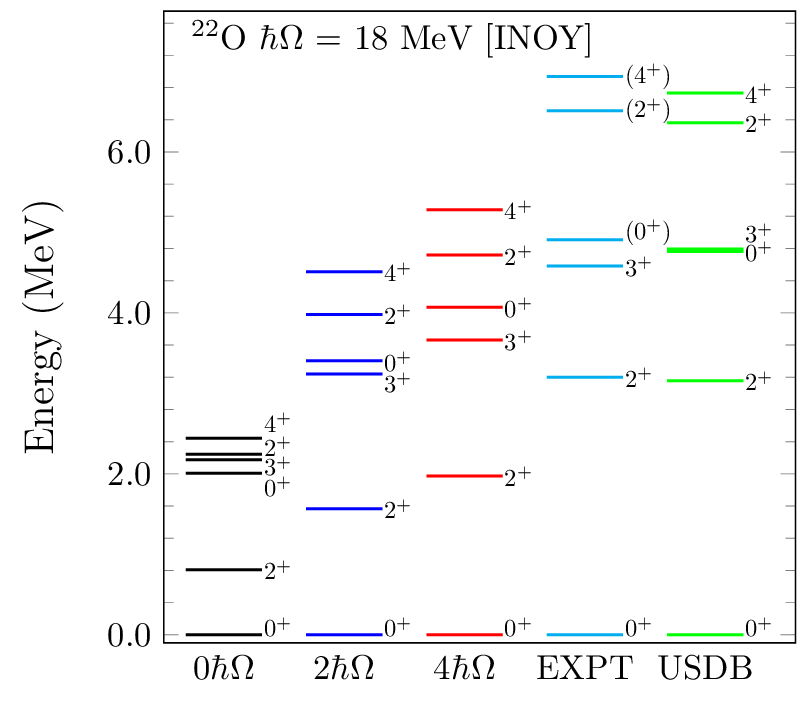}
\includegraphics[width=8cm,height=7.0cm,clip]{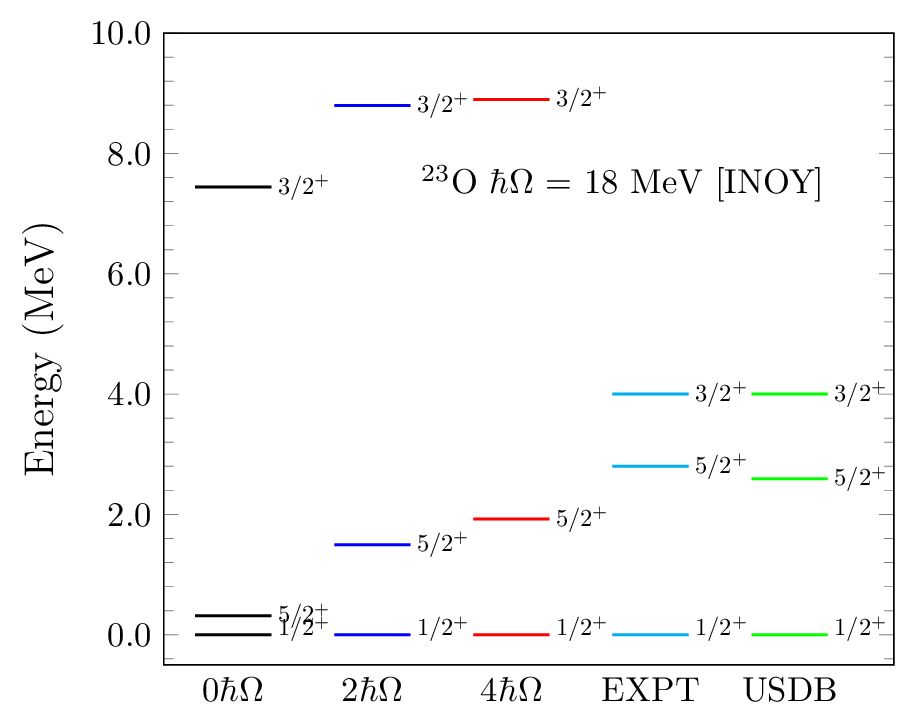}
\caption{\label{O_2123} The energy spectra of  $^{21-23}$O isotopes  with INOY and USDB interactions.}
\end{center}  
\end{figure*}

\begin{figure*}
\includegraphics[width=8cm,height=7cm,clip]{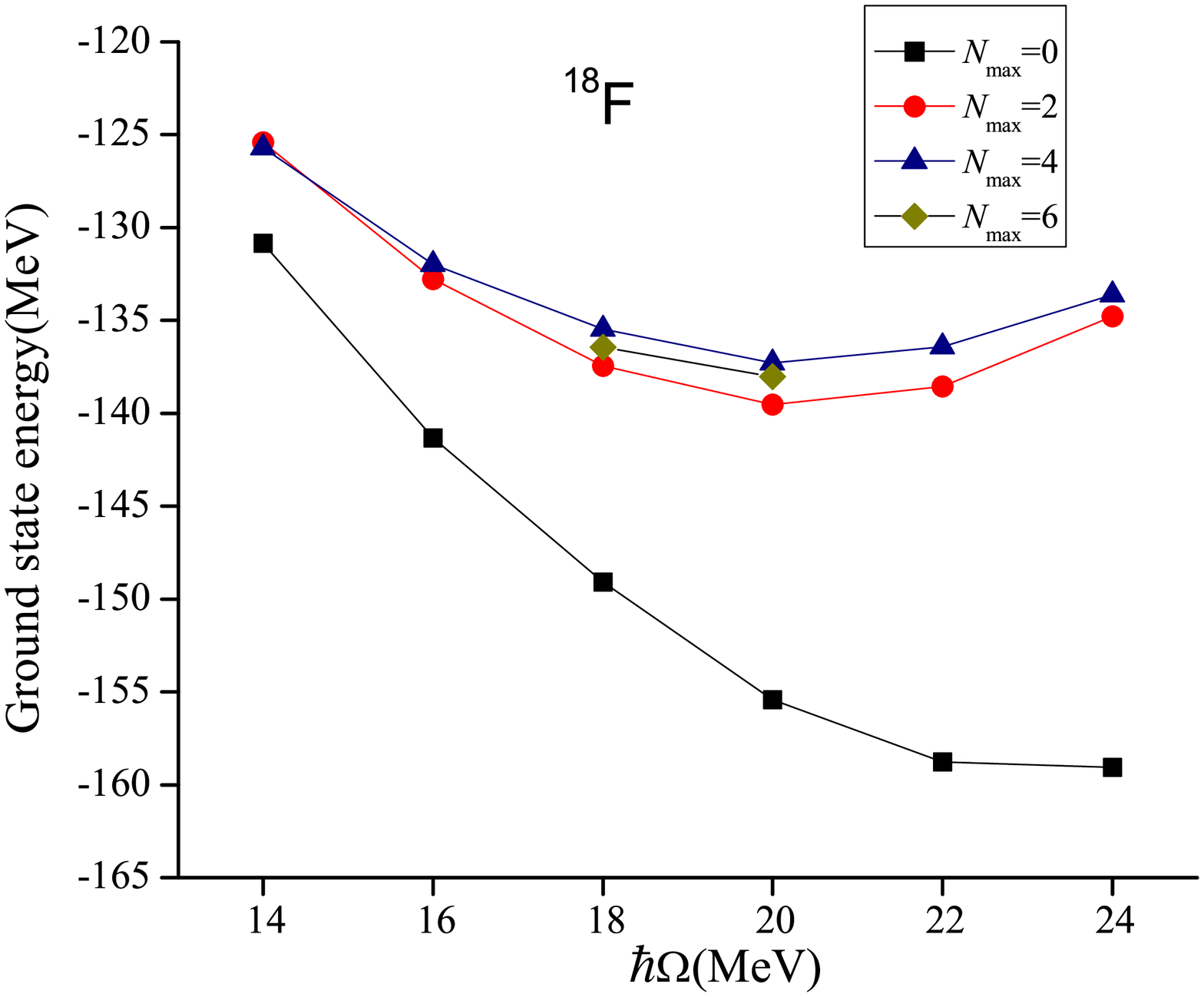}
\includegraphics[width=8cm,height=7cm,clip]{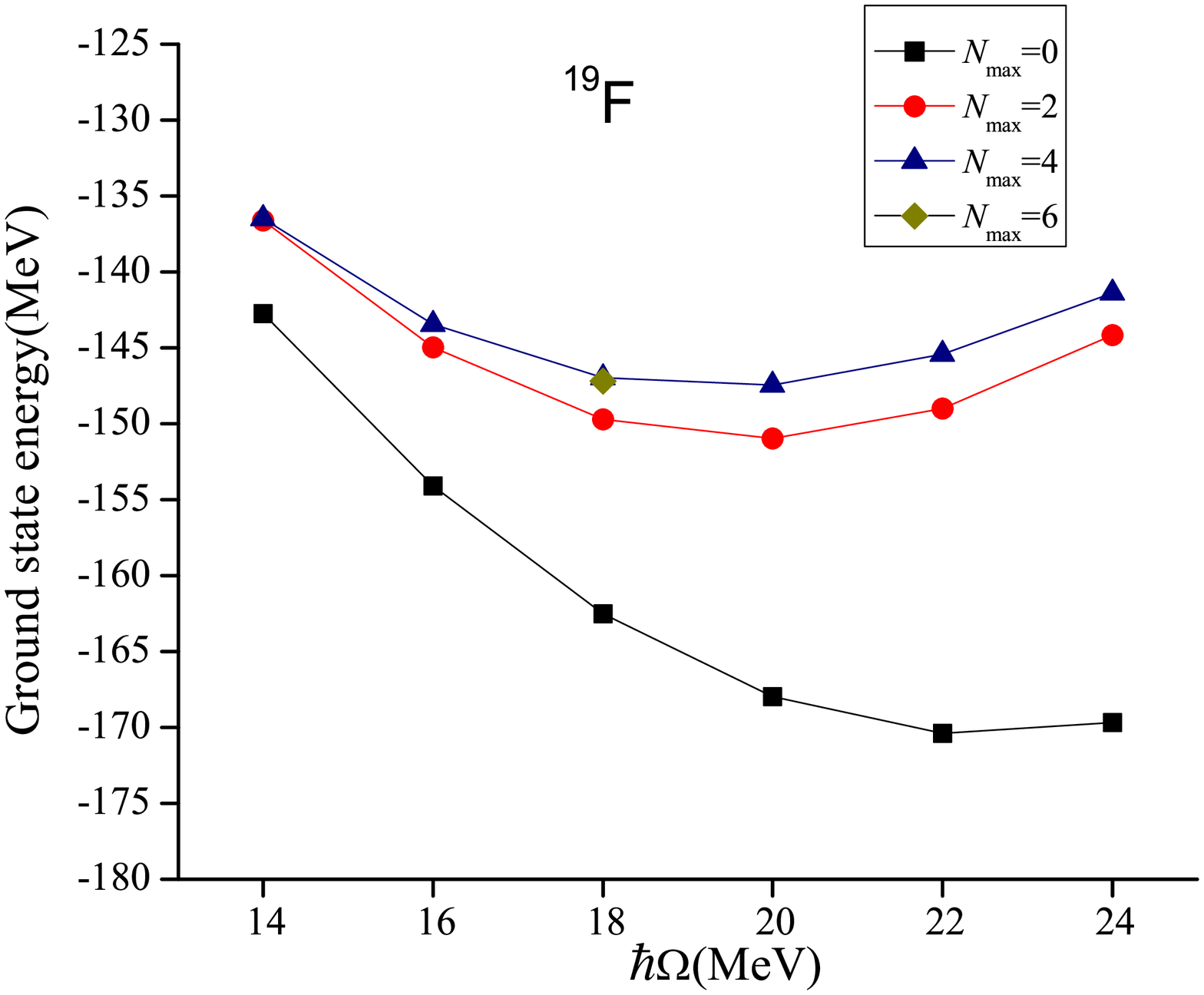}
\caption{\label{Fluorine_hw}The variation of g. s. energy with frequencies at different model spaces for $^{18-19}$F isotopes, 
similarly for other F isotopes.}   
\end{figure*}

\begin{figure*}
\includegraphics[width=16cm,height=7.5cm,clip]{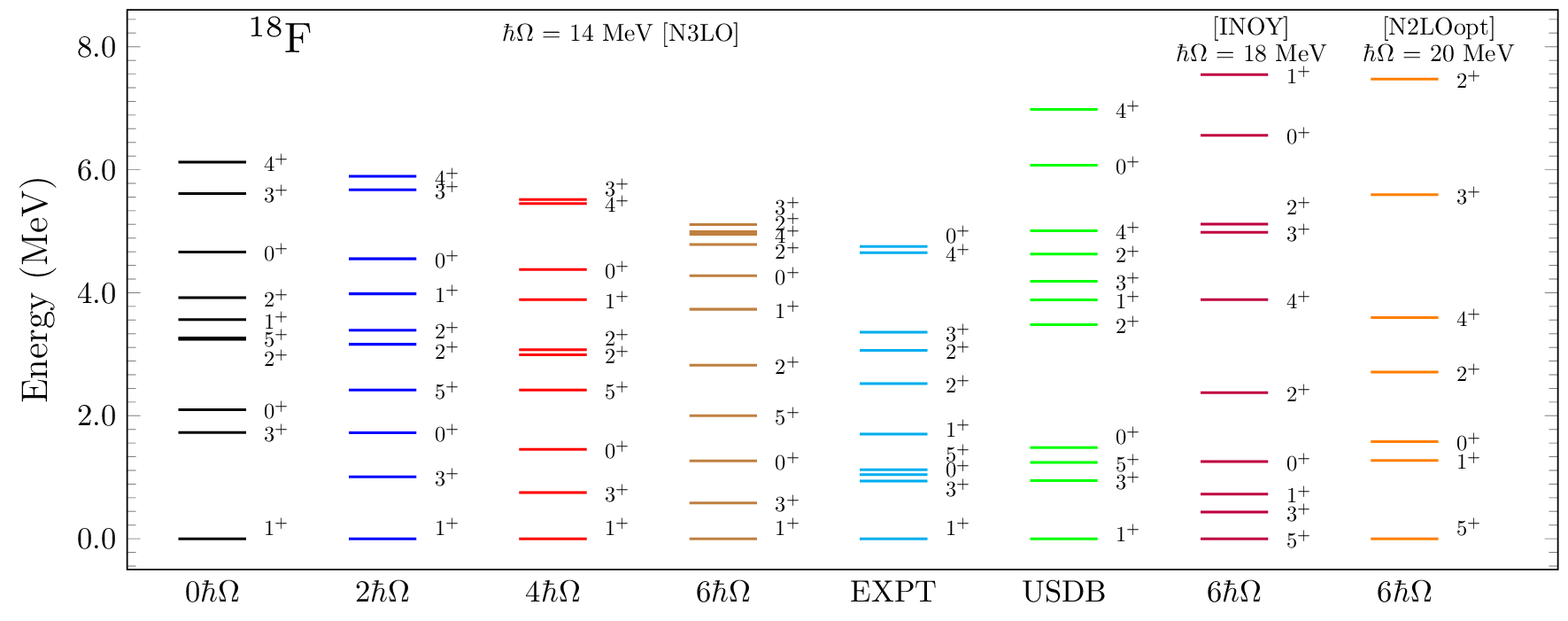}
\includegraphics[width=16cm,height=7.5cm,clip]{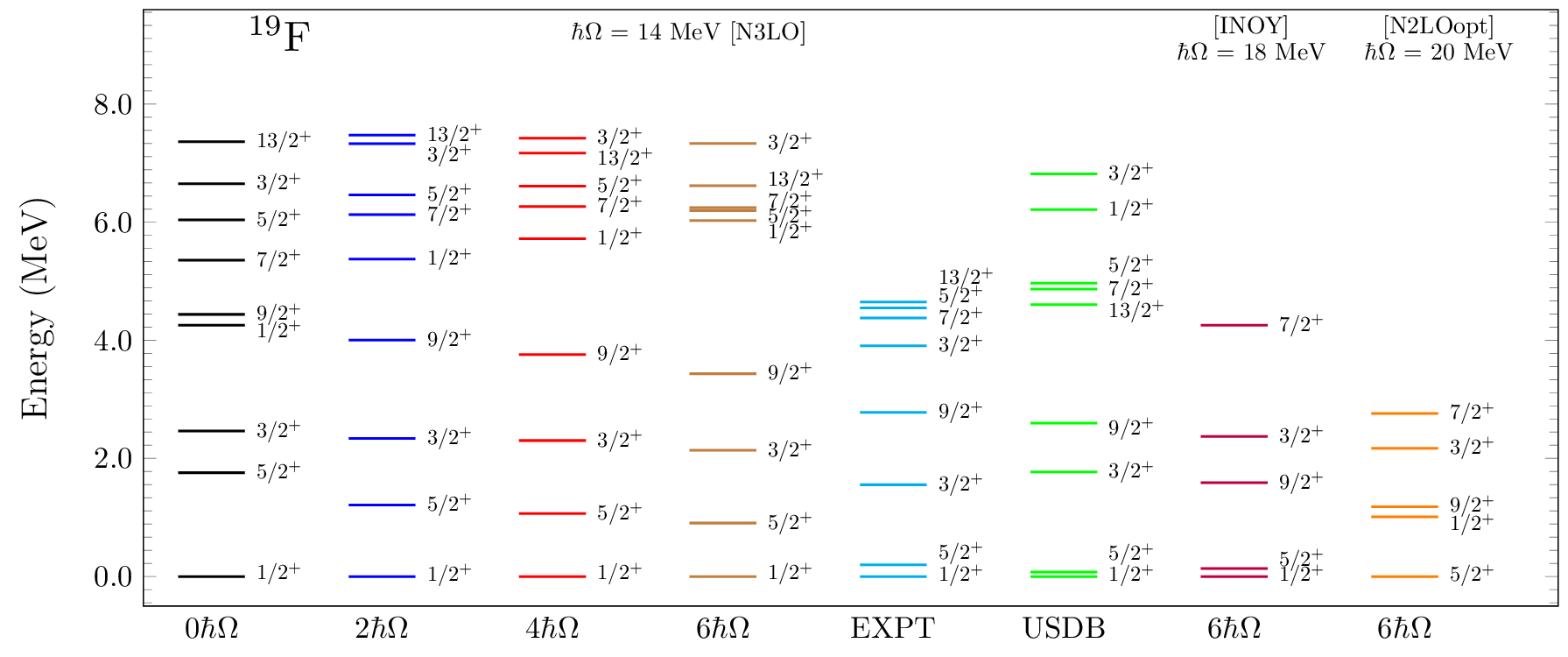}
\includegraphics[width=16cm,height=7.5cm,clip]{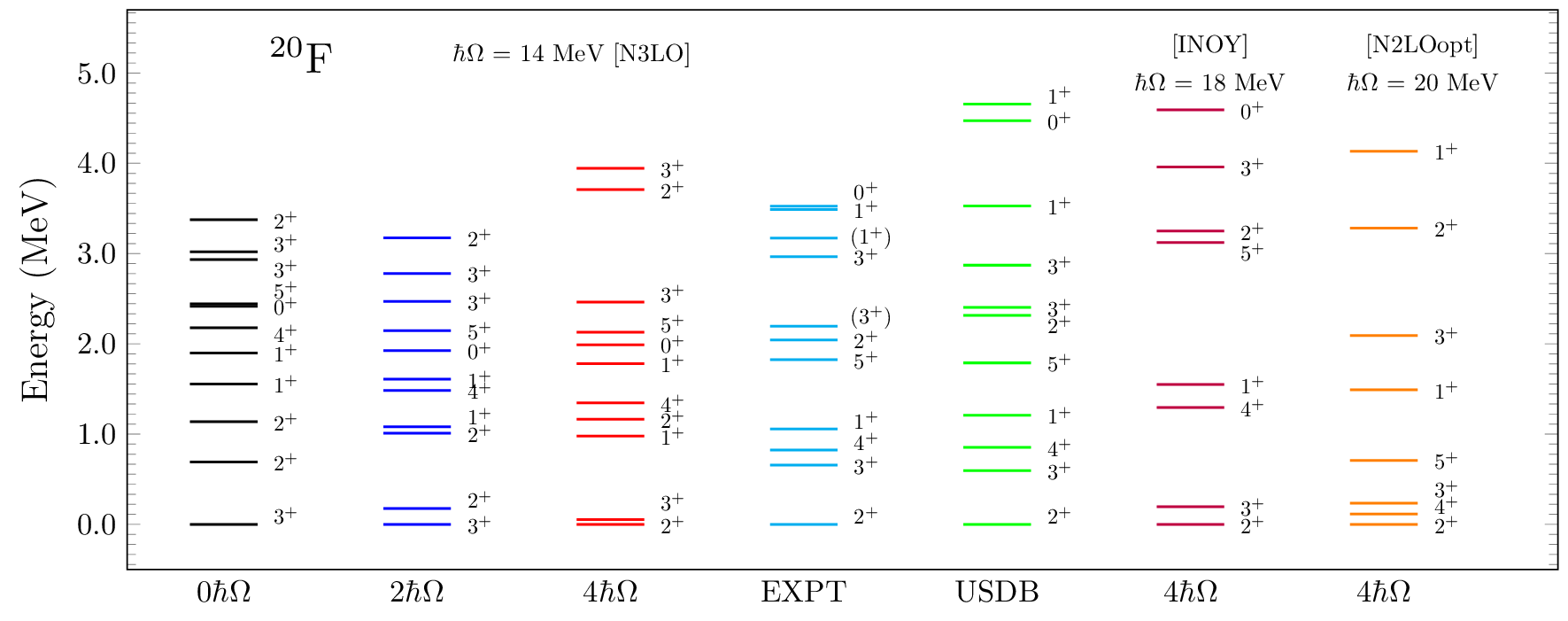}
 \caption{\label{F_1820}The energy spectra of $^{18-20}$F isotopes with N3LO,  USDB, INOY and N2LOopt interactions.}
\end{figure*}

\begin{figure*}
\includegraphics[width=16cm,height=7.5cm,clip]{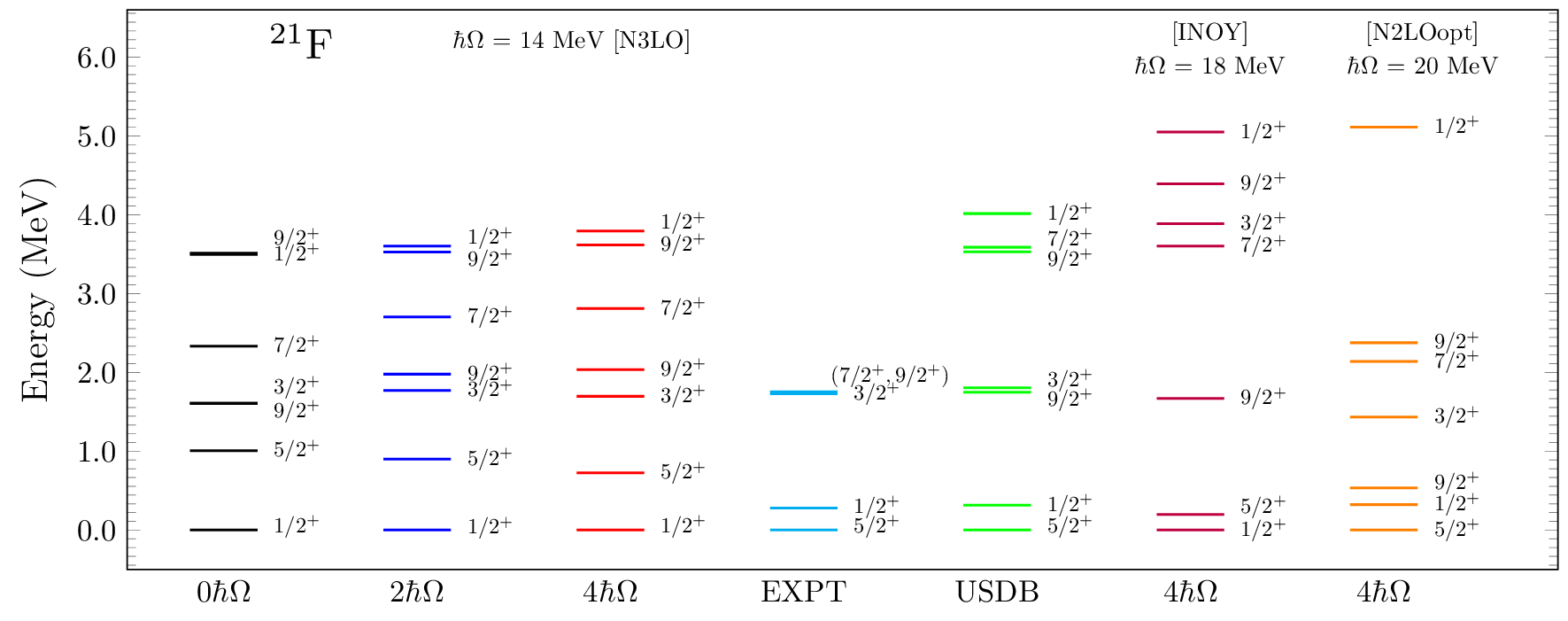}
\includegraphics[width=16cm,height=7.5cm,clip]{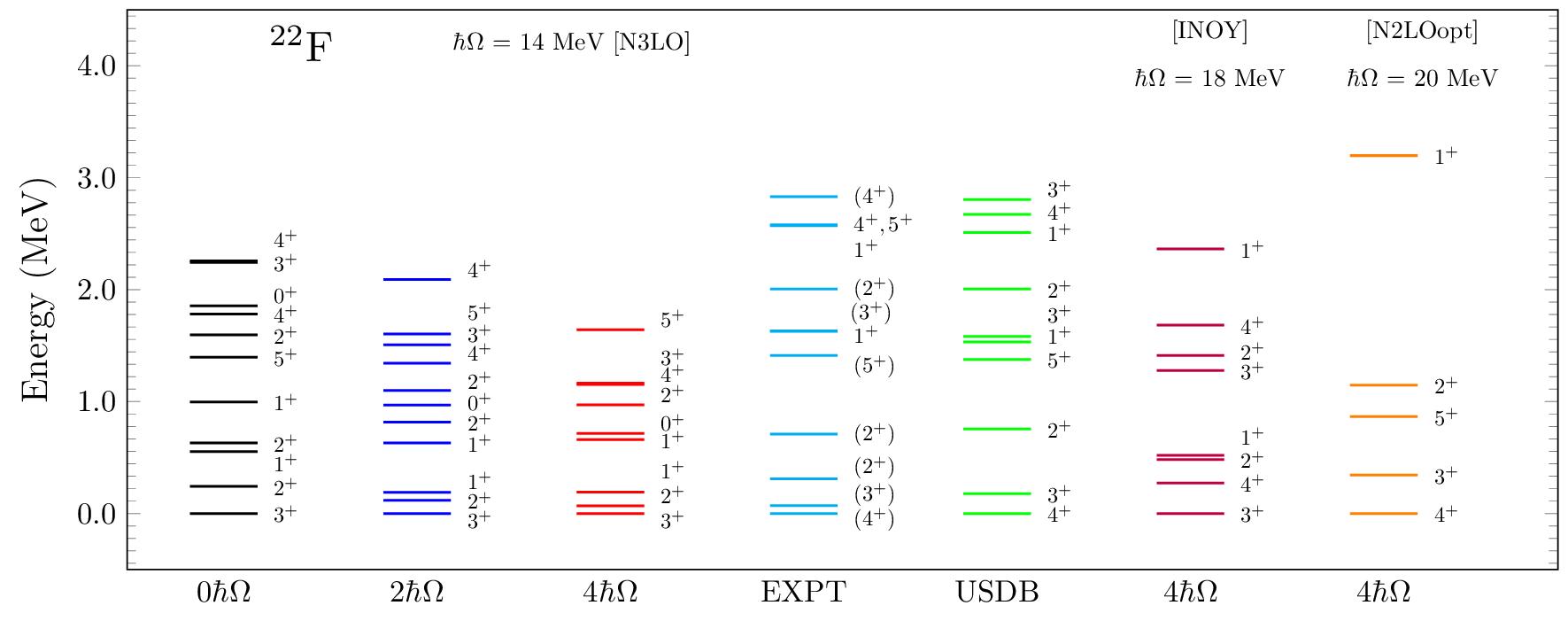}
\includegraphics[width=16cm,height=7.5cm,clip]{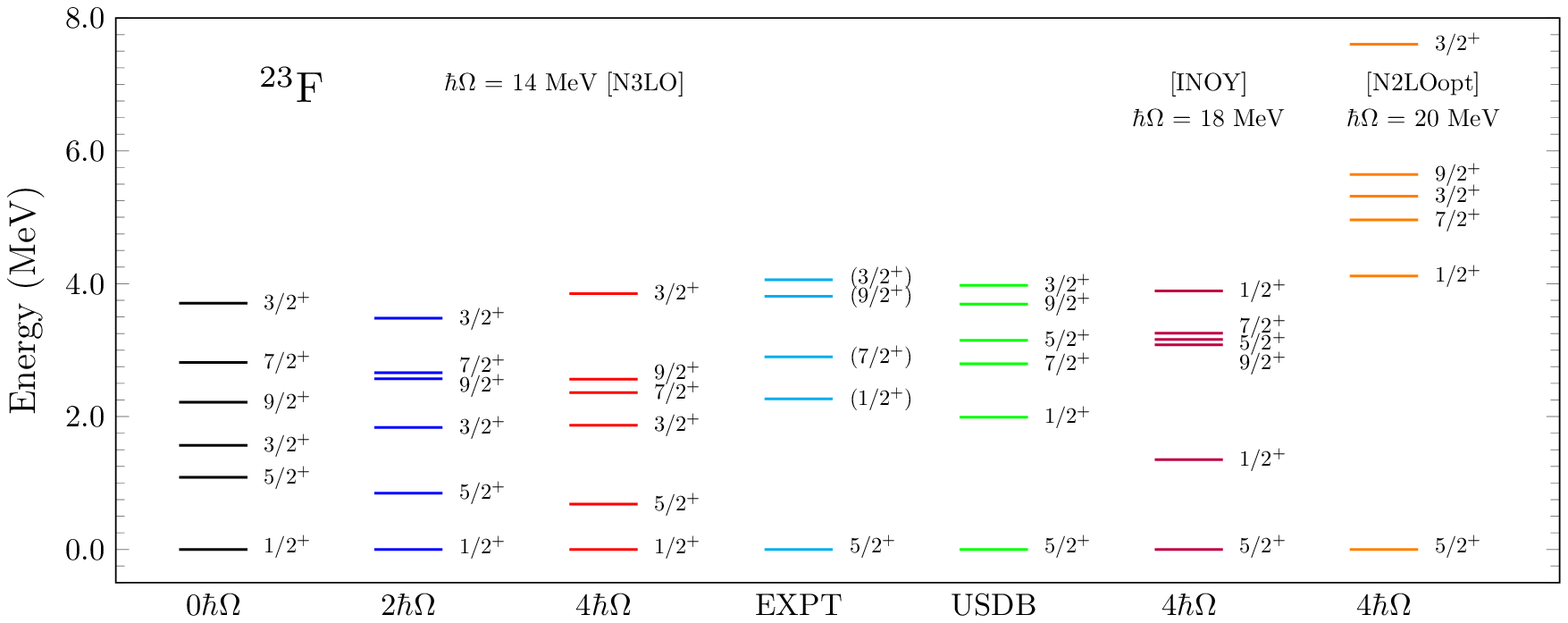}
\caption{\label{F_2123}The energy spectra of $^{21-23}$F isotopes with N3LO, USDB, INOY  and N2LOopt interactions.}   
\end{figure*}

\begin{figure*}
\includegraphics[width=16cm,height=7.5cm,clip]{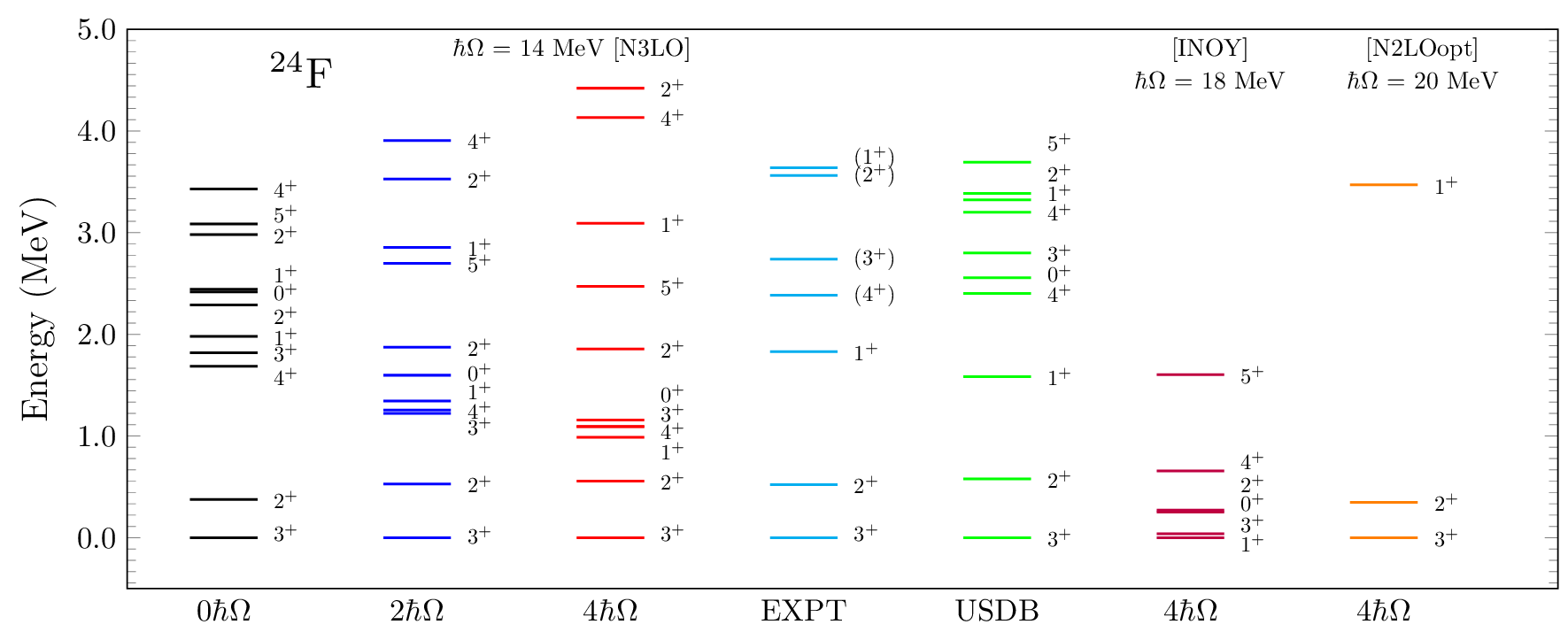}
\caption{\label{F_24}The energy spectra of $^{24}$F isotope with N3LO, USDB, INOY and N2LOopt intercations.}   
\end{figure*}

\begin{figure}
\includegraphics{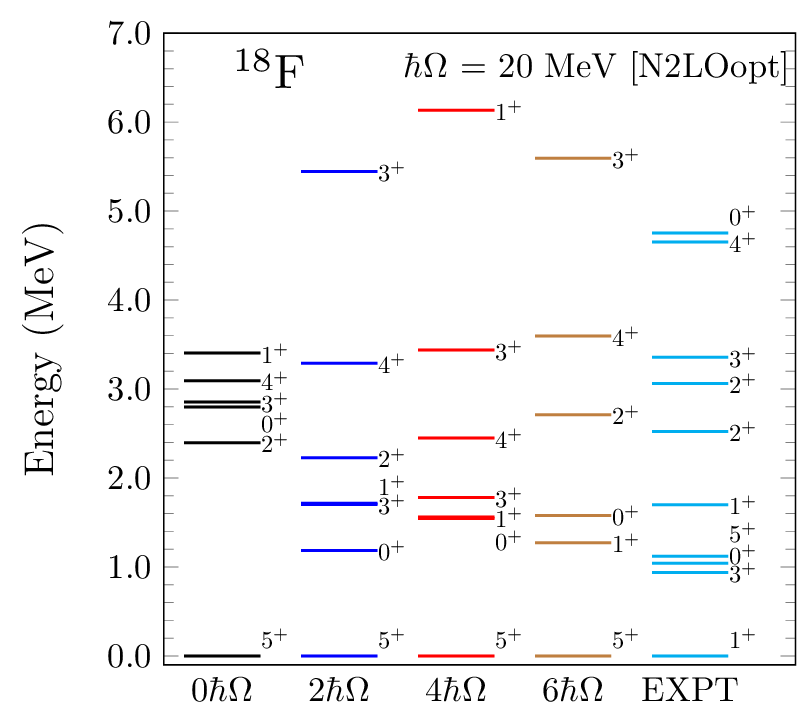}
\caption{\label{F18_N2LO}The energy spectra of $^{18}$F with N2LOopt.}   
\end{figure}


\begin{figure*}
\includegraphics[width=9.5cm,height=7cm,clip]{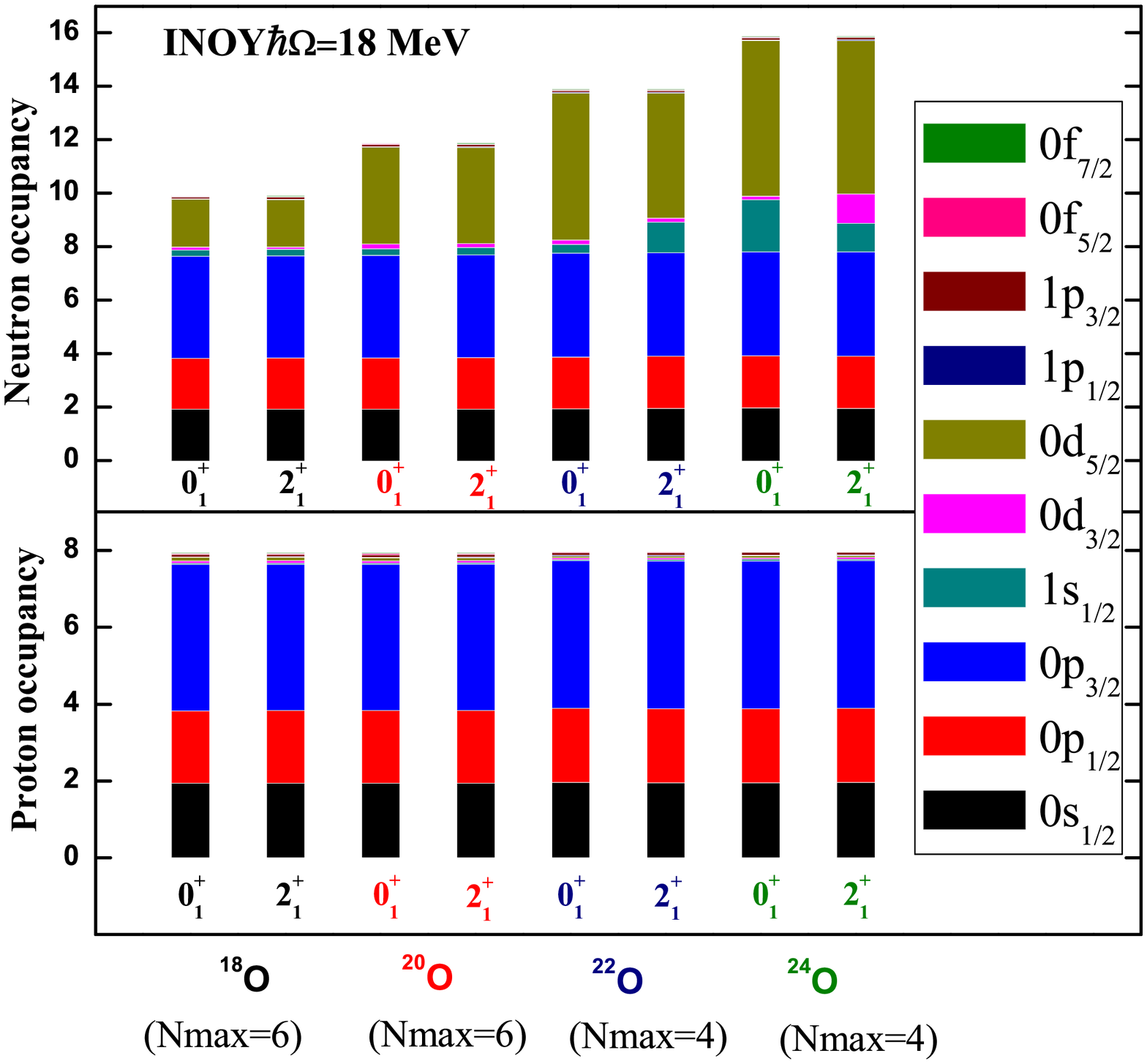}
\includegraphics[width=9.5cm,height=7cm,clip]{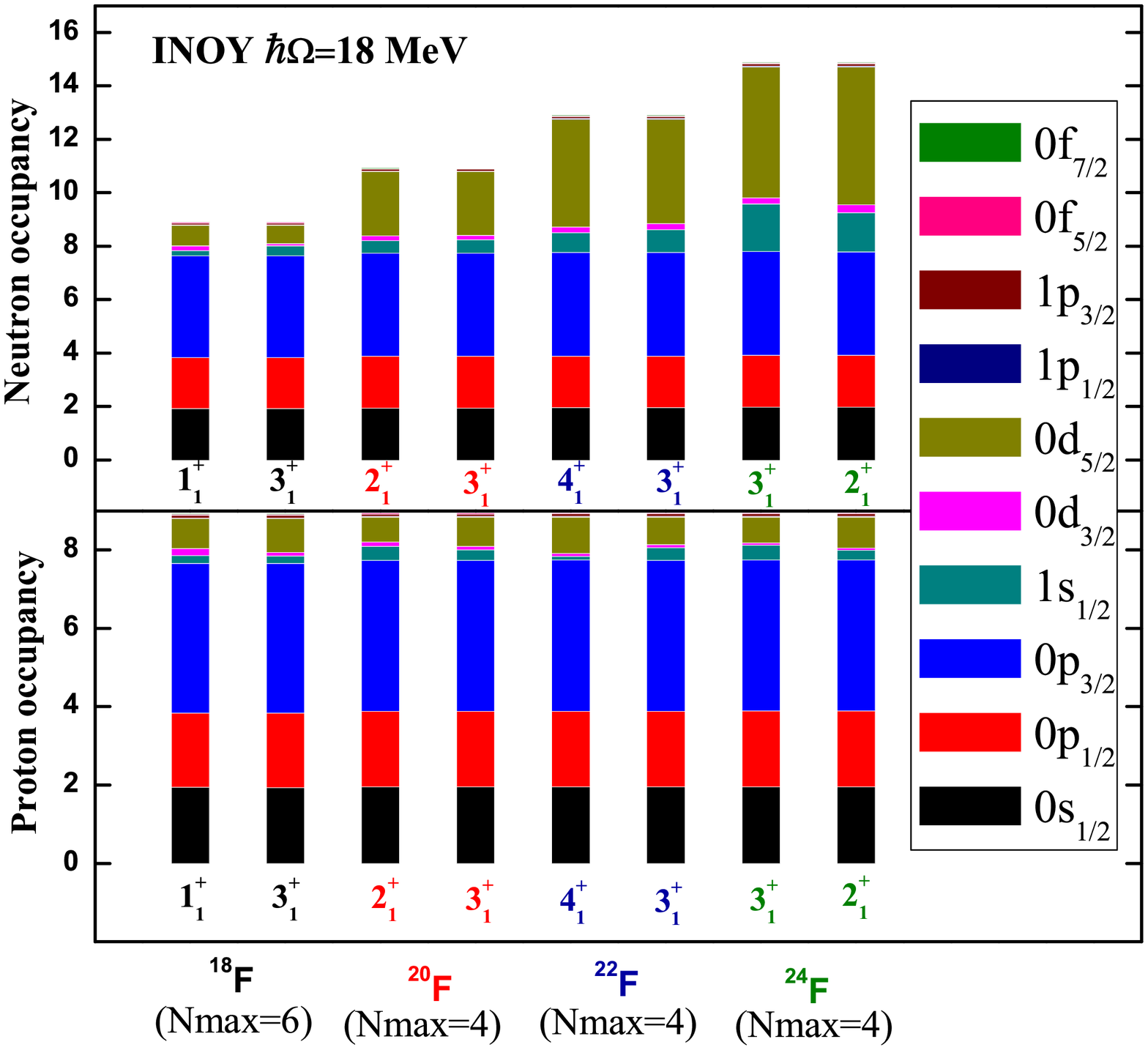}
\caption{\label{occupancy_OF}Systematics of occupation numbers for even O and F isotopes.}
\end{figure*}

\section{Results and discussions}

We have performed the NCSM calculations for oxygen ($^{18-23}$O) and fluorine chains ($^{18-24}$F). NCSM calculations are variational which depend on $\hbar\Omega$ 
and size of model space $N_{max}$. First we perform calculations with different
frequencies for a given $N_{max}$. 
We are interested in the region in which the dependence of g.s. energy on 
frequency is minimum (for largest model space).
We select that frequency for NCSM calculations. This procedure is called optimization of frequency.
 When we use this frequency, we get faster convergence than other values of frequencies.
 This is the benefit for doing optimization of frequency. When we go to higher model space, the dependence on frequency decreases.

In Figs. \ref{Oxygen_hw} and \ref{Fluorine_hw}, we have shown the variation of g.s. energies with
the harmonic oscillator frequencies and different model space sizes. We can see the g.s. energy becomes 
less dependent when we move to higher $N_{max}$. We obtain a minima at $\hbar\Omega$=20 MeV.
As we will go higher model space, we expect that this minima will shift at $\hbar\Omega$=18 MeV. 
We pick up the frequency 18 and 20 MeV to perform NCSM calculations for the energy spectra.
 We have shown the energy spectra using different interactions for oxygen 
and fluorine chains in Figs. \ref{O_1820}-\ref{O_2123}  and Figs. \ref{F_1820}-8, respectively. We have also compared  our NCSM results with
phenomenological interaction USDB.

In the case of $^{18}$O,  as we move towards the higher $N_{max}$ from 0 to 6, the calculated results also improve in comparison to lower $N_{max}$ (however,
$0_{2}^+$ is at higher energy for $N_{max}$=6  in comparison with the lower $N_{max}$).
The gap between $0_1^+$ and  $2_1^+$ increases smoothly as we move to higher $N_{max}$.
The results with $\hbar \Omega$=18 MeV are better than that of  $\hbar\Omega$=20 MeV. At $\hbar \Omega$=18 MeV,
the $2_1^+$ and $4_1^+$ states follow the same trend but the $0_2^+$ and $2_2^+$ sates are in reverse order 
in comparison to the experimental data. This reverse order is also seen with USDB.
In the case of $^{19}$O, we are not getting the correct g.s. $5/2^+$ with both the frequencies. The calculated  $5/2_2^+$ state is at higher energy.
The calculated $9/2_1^+$ state is near to the experimental data. 
The $1/2^+$ state is going far for $N_{max}$=6 in comparison with the lower $N_{max}$ results. 
In case of $^{20}$O, the low-lying spectra is in correct order below energy $\sim$2.5 MeV for $N_{max}$=6. The  $4_1^+$ state is lower in energy as compared to the experimental data.
For $^{21}$O, NCSM predicts g.s. as $5/2^+$ (although it is tentative experimentally). 
For $^{22}$O, we get slightly better results than the lighter oxygen isotopes. Here, the states with NCSM calculations are in same order as in the experimental data. The states  $0_2^+$ and $3_1^+$  are in reverse order with USDB interaction.  The INOY interaction with  $N_{max}$ =4 calculations is not able to produce the large  $N = 14$ shell gap, because $2_1^+$ is at low-excitation energy, thus higher $N_{max}$ calculations are needed.
This correlation is also reflected from the predicted $1/2_1^+$, $3/2_1^+$ states in $^{21}$O and $5/2_1^+$ in the $^{23}$O at low energy with $N_{max}$ = 4 calculations for INOY. 
For $^{23}$O, we get the correct g.s. with NCSM. The calculated $5/2^+$ state is approaching  towards  the experimental $5/2^+$ state with increasing model space size.

First time we have reported NCSM results of $^{18,19,20,21}$O isotopes for $N_{max}$ = 6. For $^{18,20}$O our results show that as we go from $N_{max}$ = 2 to 6, there is a little improvement in the result of 
$2^+$ and $4^+$ states.

In the case of fluorine isotopes ($^{18-24}$F), we show the NCSM results using three interactions N3LO, INOY and N2LOopt at $\hbar\Omega$= 14 MeV, 18 MeV and 20 MeV, respectively.

For F chain we have done calculations for $N_{max}$= 0, 2, 4, and 6 (for $^{18,19}$F)  in the case of N3LO, while for INOY and N2LOopt higher $N_{max}$  calculations only.
For $^{18}$F (see Fig. \ref{F_1820}), we can see that the NCSM calculations using N3LO interaction are better than INOY and N2LOopt interactions.
 The N3LO interaction predicts correct g.s. $1^+$ while INOY and N2LOopt fail to predict correct ground state. The order of low lying states
 up to $5_{1}^+$ is correctly reproduced with N3LO interaction while other three interactions are not able to reproduce correctly. 
 The USDB interaction predict $5_{1}^+$ lower in comparison to $0_{1}^+$. With N3LO, the calculated $0_{1}^+$ state show rapid convergence with increasing $N_{max}$. The $N_{max}$ = 6 results are close to the experimental data. The calculated $0_{2}^+$ state with N3LO is further compressed with increasing $N_{max}$.
The calculated value of  $3_{1}^+$ corresponding to $N_{max}$=6 is compressed in comparison to the experimental data.
The difference between calculated and  experimental excitation energy of $1_{2}^+$ state is very large.
To see the N2LOopt results more systematically with different $N_{max}$,  we have shown corresponding results for $^{18}$F in the Fig. \ref{F18_N2LO}.
 The predicted $1^+$ state is at high energy ($>$ 1 MeV) even with $N_{max}$ = 6 calculation.

In case of $^{19}$F, the $5/2_{1}^+$ state is close to the experimental data for INOY interaction 
and the order of $3/2_{1}^+$ and $9/2_{1}^+$ is reverse, while for the N3LO interaction the  $5/2_{1}^+$ state is at high in energy in comparison to the 
experimental data and it predicts the correct order for $3/2_{1}^+$ and $9/2_{1}^+$ states.
 With increasing $N_{max}$ the excited states $5/2_{1}^+$-$3/2_{1}^+$-$9/2_{1}^+$ predicted by N3LO interaction 
are showing fast convergence, while gap between $9/2_{1}^+$-$1/2_{2}^+$ is 
increases as $N_{max}$ reaches from 0 to 6.
N2LOopt interaction fails to reproduce the correct g.s. $1/2^+$.
As we move from $^{18}$F to $^{19}$F, the INOY interaction results are much better.

From $^{20}$F onwards, it is not possible to perform NCSM calculation for $N_{max}$=6, thus we have reported results up to  $N_{max}$=4 only. 
For $^{20}$F, all the three interactions reproduce the correct g.s. $2^+$. The NCSM results using INOY interaction are better in comparison to  N3LO and N2LOopt. 
 The calculated $1_1^+$ state with N3LO and USDB interactions is showing reasonable agreement with the experimental data.

 In the case of $^{21}$F, there is an inversion of the lowest $1/2^+$ and $5/2^+$ states in the spectra obtained from N3LO and INOY interactions in comparison with the experimental data. We need higher Nmax calculations for these interactions because $5/2^+$ state is converging very fast.
In the case of $^{21}$F and $^{22}$F the N2LOopt interaction reproduces the correct g.s. $5/2^+$ and $4^+$, respectively. For $^{21}$F, the NCSM results show considerable gap between $9/2_{1}^+$ and $7/2_{1}^+$
states in comparison to the experimental data. 
 Overall the N3LO  prediction is not good for $^{22}$F. It is clear that even if we do large  $N_{max}$
calculations our results will not be improved. For this isotope, the INOY interaction is also not able
to predict correct g.s.

In the case of $^{23}$F, all the interactions (except N3LO) predict correct g.s. as $5/2^+$.  The $5/2^+$ state is converging very fast with INOY. Thus it is expected that if we do large $N_{max}$ calculation then this interaction will be able to predict correct g.s.

 In the case of $^{24}$F, except INOY, all other interactions  predict correct g.s. as $3^+$. Thus, higher $N_{max}$ calculations are needed for INOY. The first excited  $2^+$ state is 
predicted correctly with N3LO. The N3LO,
predicts $1_{1}^+$ and $2_2^+$ states at much lower energy in comparison with the experimental data. The $1^+$ and $3^+$ states are very close with INOY interaction. If we do higher $N_{max}$ calculation for INOY then we get correctly 
 g.s. as $3^+$.
The N2LOopt interaction gives $1_{1}^+$ state at very high energy.
In the case of all fluorine isotopes we see the energy levels are more spread using INOY interaction in comparison with N3LO interaction. This is because of the strong ${\bf \Vec{l}.\Vec{s}}$ coupling in the INOY interaction.


Overall, for better description of energy spectra in O and F chains we need to go to higher $N_{max}$ calculations. 
Fig. \ref{occupancy_OF} shows the occupancies for the ground  and first excited states for the even oxygen and fluorine chains, respectively. We have shown the occupation numbers up to
$fp$ shell. Above the $fp$ shell the occupation numbers are very small and very hard to visualize (in the present figure).
So, we have not shown the orbitals beyond $fp$ shell. 
 The occupancy of $\nu d_{5/2}$ orbital is increasing from $^{18}$O to $^{24}$O. For $^{24}$O, the occupancy of the $\nu d_{3/2}$
orbital show significant increase from $0_1^+$ to $2_1^+$ state, while $\nu s_{1/2}$ occupancy is decreasing.

\begin{figure}[h]
\includegraphics[width=8.8cm,height=7cm,clip]{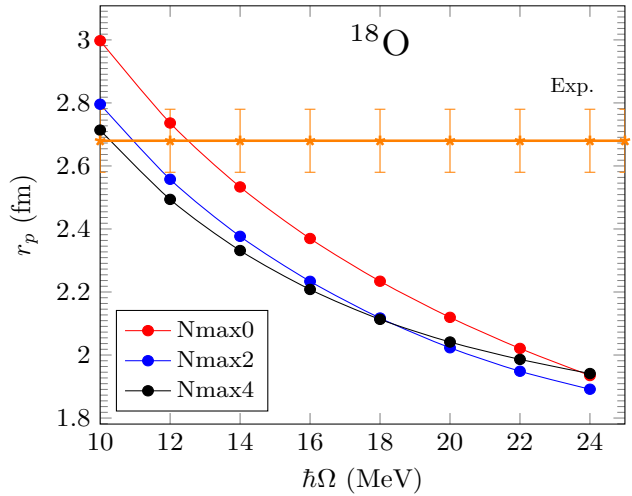}
\caption{\label{Radii_O}
$^{18}$O point-proton radius as a function of $\hbar\Omega$ obtained with INOY interaction in NCSM calculations for $N_{max}$=0, 2, and 4. The experimental value with uncertainties \cite{Lapoux} is also shown.  }
\end{figure}

In the Fig. \ref{Radii_O} we have shown $^{18}$O point-proton radius as a function of $\hbar\Omega$ obtained with INOY interaction. The crossing point of the INOY curves suggests that the $r_{p}$ $\approx$ 2.12 $fm$  at 18 $\hbar\Omega$, while corresponding experimental value is 2.68(10) $fm$. 

\section{Location of drip line in oxygen isotopes}

\begin{figure*}
\includegraphics[width=8.8cm,height=7cm,clip]{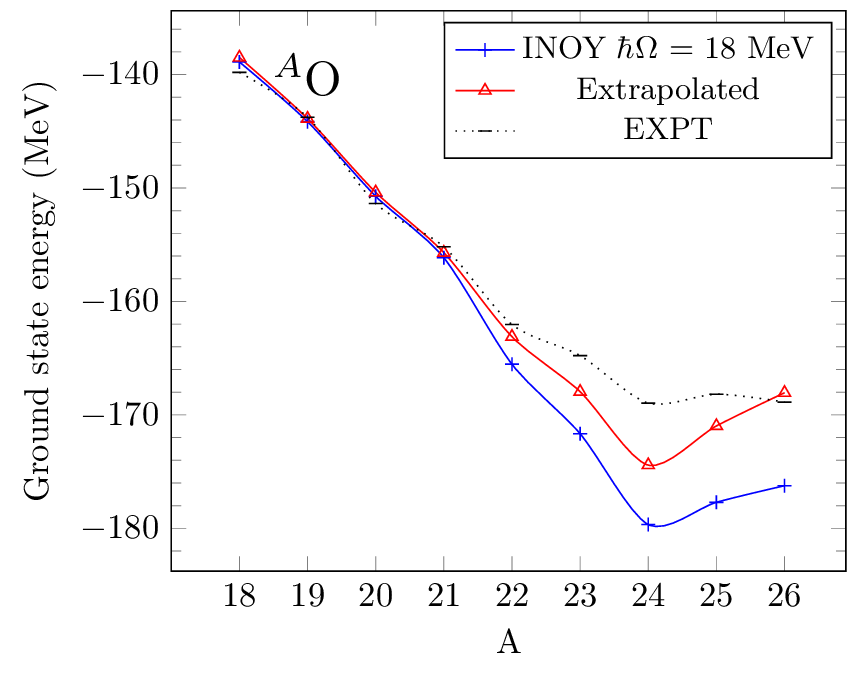}
\includegraphics[width=8.8cm,height=7cm,clip]{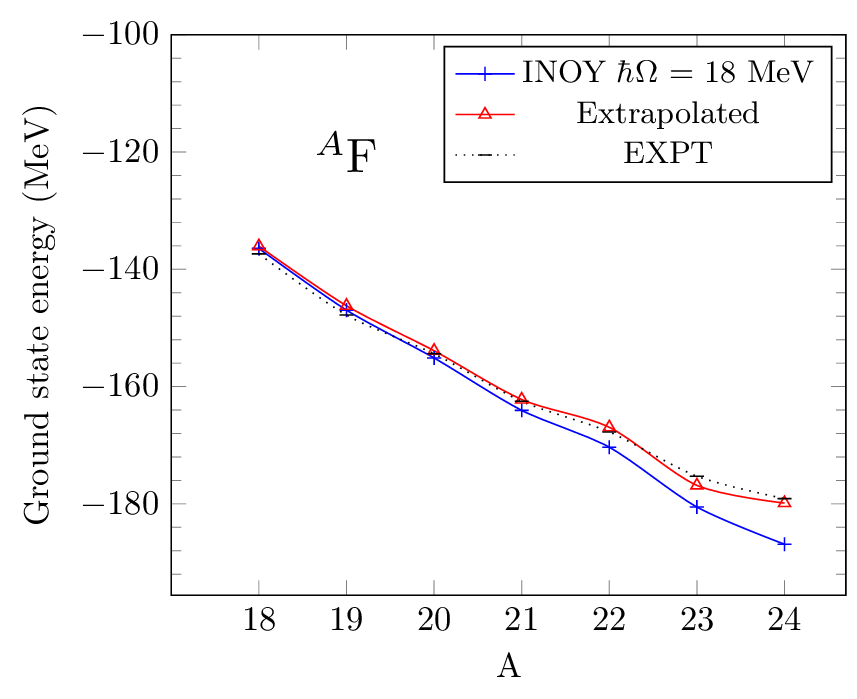}
\caption{\label{GS_OF}
Comparison of experimental, calculated (corresponding to $\hbar\Omega$ = 18 MeV) and extrapolated g.s. energies of O and F isotopes with INOY interaction.
For $^{18-21}$O and $^{18,19}$F the g.s. energy is for $N_{max}=6$ and for rest isotopes with $N_{max}=4$. }
\end{figure*}

As we know from the previous work in oxygen chain, the 3N forces is needed to reproduce the drip line at $^{24}$O  \cite{Otsuka, Hagen, Cipollone}. 
The calculations are previously done \cite{Otsuka, Hagen, Cipollone} using chiral $NN$, $NN+NNN$-induced and $NN+NNN$-full interactions. 
The anomalous behaviour  in oxygen isotopes can be explained after adding 3N forces only. The $NN$ interaction shows
the drip line at $^{28}$O \cite{Otsuka,Hergert,Roth}. In the present work, we have applied INOY $NN$ interaction
which have the effect of three body forces in terms of short range and nonlocality character present in it and it 
also reproduces the correct binding energy of triton.
In the upper panel of Fig. \ref{GS_OF}, we have shown the g.s. energies for oxygen isotopes using INOY at $\hbar\Omega$=18 MeV. The g.s. energy decreases as we reaches to $^{24}$O but we see a kink after this. This shows the drip line in oxygen isotopes at $^{24}$O. 
The g.s. energy for fluorine isotopes is shown in the lower panel of Fig. \ref{GS_OF}.
Using INOY interaction the g.s energies are quite good up to $^{21}$F. 
We have extrapolated the g.s. energy using an exponential fitting function $E_{g.s.}(N_{max})= a~exp(-bN_{max}) + E_{g.s.}(\infty)$ with 
 $E_{g.s.}(\infty)$ the value of g.s. energy at $N_{max}$ $\rightarrow$ $\infty$  for $^{18-21}$O using 
 $N_{max}$ = 0, 2, 4 and 6 results, while beyond $^{21}$O $N_{max}$ = 0, 2, and 4 results only.

The location of drip line for O and F chain using hybrid $ab~initio$ approach is reported by Tichai et al in Ref. \cite{Roth}. In this work \cite{Roth} low-lying spectra of $^{18,19,20}$O isotopes up to $N_{max}$ = 6 are also reported.
Otsuka et al, previously reported  first microscopic explanation of the oxygen anomaly based on three-nucleon forces 
in Ref. \cite{Otsuka}.


\section{Conclusions} 
In the present work first time we have reported no core shell model results of 
 oxygen and fluorine chain using INOY $NN$ interaction.
 Also first time we have  done NCSM calculations for larger $N_{max}$ (up to $N_{max}$ =6 for some isotopes)
for O and F chain.  The NCSM study with INOY $NN$ interaction is very important because it gives the effect of three body forces without adding three body forces explicitly.   
We have reported energy spectra for positive parity states and 
neutron and proton occupancies for the maximum reached $N_{max}$ in our calculations.
The NCSM calculations with INOY interaction predict correctly the drip line at $^{24}$O for oxygen chain.
The large $N = 14$ shell gap is not reproduced with INOY interaction. However, if 
we perform higher $N_{max}$ calculations with INOY interaction then it is possible to get large $N = 14$ shell gap.
 The INOY interaction is unable to correctly reproduce the experimental data for $^{18}$F isotope, however as we move to
heavier neutron rich F isotopes the results are in a reasonable agreement with the experimental data
in comparison to N3LO and N2LOopt interactions.



\section*{Acknowledgment:}

We would like to thank Prof. Petr Navr\'atil for providing us his NN effective interaction code and for his valuable comments and suggestions from the beginning of this project. Thanks are also due to Prof. Christian Forss\'en for pAntoine and for several 
discussions. AS acknowledges financial support from MHRD (Govt. of India) for her Ph.D. thesis work. We acknowledge Oak computational facility at TRIUMF for No-Core Shell Model calculations. P.C.S. acknowledges the hospitality extended to him during his stays at TRIUMF in the summer of 2016 and 2018.


\end{document}